\documentclass[twocolumn,showpacs,preprintnumbers,amsmath,amssymb,superscriptaddress]{revtex4}
\usepackage{graphicx}

     \let\d=\delta

\def\CC{{\cal C}}

\def\atanh{{\rm atanh}}

\def\to{\rightarrow}

\newcommand{\beq}{\begin{equation}}
\newcommand{\eeq}{\end{equation}}

\begin{document}

\title{Dynamics and termination cost of spatially coupled mean-field models}

\author{Francesco Caltagirone}
\affiliation{Institut de Physique Th\'{e}orique, CEA, CNRS-URA 2306, F-91191,
Gif-sur-Yvette, France}
\author{Silvio Franz}
\affiliation{Laboratoire de Physique Th\'{e}orique et Mod\`{e}les Statistiques, 
Universit\'{e} Paris-Sud 11 and CNRS UMR8626, Bt. 100, 91405 Orsay, France}
\author{Richard G.~Morris}
\affiliation{Institut de Physique Th\'{e}orique, CEA, CNRS-URA 2306, F-91191, 
Gif-sur-Yvette, France}
\author{Lenka Zdeborov\'{a}}
\affiliation{Institut de Physique Th\'{e}orique, CEA, CNRS-URA 2306, F-91191, 
Gif-sur-Yvette, France}

\begin{abstract}
This work is motivated by recent progress in information theory and signal
processing where the so-called `spatially coupled' design of systems leads
to considerably better performance. We address relevant open questions about
spatially coupled systems through the study of a simple Ising model.  In 
particular, we consider a chain of Curie-Weiss models that are coupled by
interactions up to a certain range. Indeed, it is well known that the pure 
(uncoupled) Curie-Weiss model undergoes a first order phase transition driven 
by the magnetic field, and furthermore, in the spinodal region such systems are 
unable to reach equilibrium in sub-exponential time if initialized in the 
metastable state.  By contrast, the spatially coupled system is, instead, able 
to reach the equilibrium even when initialized to the metastable state. 
The equilibrium phase
propagates along the chain in the form of a travelling wave.  
Here we study the speed of the wave-front and the so-called `termination 
cost'--- \textit{i.e.}, the conditions necessary for the 
propagation to occur. We reach several interesting conclusions about 
optimization of the speed and the cost.
\end{abstract}

\pacs{05.70.Fh, 89.20.Ff, 02.50.Tt} 

\maketitle

\section{Introduction}
\label{sec:intro}

Many questions of interest in modern science can be formulated as
inference problems--- where there is a set of variables (the signal) on which 
we are only able to perform some kind of partial, aggregate, or incomplete 
observations; the goal being to \textit{infer} the values of the variables 
based on the indirect information contained in the measurements.  In most 
cases, this amounts to devising both a measurement protocol and a corresponding 
algorithm for reconstructing the underlying variables.  Two examples of 
problems
that fall into this category are the following:
\paragraph*{Compressed sensing:} It is well known that most signals of interest 
are compressible, however, the compression process is typically only carried 
out once the signal is known, or has been measured.  This must be contrasted 
with the fact that, in many applications, (\textit{e.g.}, medical imaging using 
MRI or tomography) it is desirable to reduce the measurement time as much as 
possible (to reduce costs, radiation exposure \textit{etc.}).  This apparent 
conflict leads to the idea of compressed sensing \cite{Donoho:06}: a signal processing method 
where data is sampled in compressed form and then the underlying signal is 
reconstructed algorithmically.
\paragraph*{Error correcting codes:}  In telephone or satellite communication, 
a signal is typically sent over a noisy channel. One may ask: is there a way to 
encode the signal in a redundant way such that it can still be reconstructed 
without errors, even after the noisy transmission?  The goal of an error 
correcting code is to optimize this redundancy rate and still to be able to 
perform exact error correction, for a review see e.g. \cite{RichardsonUrbanke08}.

Inference problems can be formulated as problems of statistical mechanics at 
finite temperature, see \cite{MezardMontanari09,KrzakalaPRX2012}.  The 
variables play the role of spins, the constraints resulting from measurements 
correspond to interactions between spins that are summarized in the Hamiltonian 
of the corresponding statistical mechanical model. The log-likelihood in the 
inference problem is therefore the negative free energy in statistical 
mechanics.  A lot of measurements, or a high signal-to-noise ratio, lead to 
interactions favoring one specific spin configuration, which corresponds to the 
original and correct signal. A decrease of the signal-to-noise ratio or of the 
number of measurements modify the free-energy landscape by giving more weight 
to spurious configurations corresponding to metastable states that start to 
compete with the low free-energy favored state and, hence, make the inference 
of the original signal difficult. An optimal algorithmic way to reconstruct 
(infer) the signal requires the computation of marginal probabilities of the 
corresponding posterior probability, in the statistical physics language this 
means computing local magnetizations of the corresponding Boltzmann 
distribution. This is computationally very difficult task and both in inference 
and in physics the most basic and popular approximations involve the mean-field 
approach (known as variational Bayes method in inference) or a Monte Carlo 
simulation (known as Gibbs sampling in inference).

In the two above examples of compressed sensing and error correcting codes (and 
many other practically important cases) the Hamiltonian of the problem can be 
designed by the engineer in order to achieve best possible performance. For 
example, in the error correcting codes the aim is to construct codes that 
contain only as much redundant information as is absolutely necessary (as 
specified famously by Shannon \cite{shannon2001mathematical}) in order to be 
able to perform the error correction. In compressed sensing, this corresponds 
to using only as many measurements as the number of elements in the compressed 
signal. It turns out that the encoding/measurement protocols able to achieve 
this optimality often correspond to random Hamiltonians, for which mean field 
solutions are exact \cite{MezardMontanari09,KrzakalaPRX2012}.  This is to be 
contrasted with statistical physics, where mean-field theory is typically a 
first tool used in order to eventually understand the behavior of more 
complicated systems.  In inference, the models of primary interest often \textit{are} 
the mean field ones--- with Hamiltonian being defined on a random graph 
(corresponding to the Bethe approximation) or on a fully connected lattice 
(corresponding to the Curie-Weiss approximation).

In the limit of large system sizes, and for a given design of the Hamiltonian, 
the best achievable performance is characterized by a phase transition that 
separates a region where inference is possible from a region where it is not.  
Just as in physics, depending on the Hamiltonian, this phase transition can be 
of first (discontinuous) or of second (continuous) order. Problems where the
transition is of first order (\textit{e.g.}, the two problems above) are much 
more algorithmically challenging for the following reason: in mean field 
systems, a first order phase transition is associated with a well defined 
spinodal region of exponentially-long living metastability. This metastability 
is due to existence of a local optimum in the posterior likelihood that is iteratively 
extremized by the inference algorithm, whereas optimal inference corresponds to 
finding the global optimum.  In physics the global optimum corresponds to the 
equilibrium state and the local optimum to a metastable state. In the context 
of inference problems the presence of a first order phase transition implies 
that the region where inference is possible divides into two parts--- the
spinodal part and the rest. In the spinodal region there exists a metastable
phase corresponding to unsuccessful inference from which it would take an
exponentially long (in the size of the system) time to reach the equilibrium 
(\textit{i.e.}, successful inference). Hence in inference problems where a 
first order phase transition appears, the corresponding spinodal line poses a 
barrier for algorithmic solution in polynomial time. Such an algorithmic 
barrier has now been identified in many different inference problems
including the compressed sensing and error correcting codes.

It is well known in statistical physics that the exponentially-long-living 
metastable states only exist in mean-field systems. In any finite dimensional 
system where the surface of a compact droplet is always smaller than its volume 
if the droplet is sufficiently large, the system escapes from the metastable 
state in a time that is polynomially proportional (often linearly) to the 
system size. This is the concept of nucleation that is well studied in physical 
sciences.

A natural question is therefore: is there a way to induce nucleation in the 
above inference problems? Recall that the corresponding Hamiltonian needs to be 
locally mean field-like in order to achieve the best possible performance.  
Hence the question is how to combine the required mean-field nature (achieving 
optimal performance) and finite dimensional nature in order to induce 
nucleation and escape from the undesirable metastable state. This can be 
achieved by a concept called `spatial coupling' in which one designs the 
Hamiltonian by taking independent copies of the mean field system and coupling 
them to close-neighbors along a one-dimensional chain. The idea then is that 
for a small part of the system called the seed (placed usually at the beginning 
of the chain) the parameters are such that successful inference is achieved in 
that part. When the seed is sufficiently large and strong, the interactions 
along the chain ensure that successful inference is also achieved for its 
neighbors, and so on.  The physical principle is the same as for a 
supercritical nucleus to start to grow during nucleation. Typically, successful 
inference is characterized by a travelling wave-like phenomenon, as the 
accurate reconstruction starts in the seed and travels `along' the 
chain.

The idea of using spatial coupling in order to improve performance in
inference problems goes back to the so called `convolutional low
density parity check codes' 
\cite{FelstromZigangirov99,lentmaier2005terminated}.  The proof of
so-called `threshold saturation', \textit{i.e.}, that the optimal inference is 
achievable, is due to 
\cite{lentmaier2010iterative,KudekarRichardson10,kudekar2012spatially}.  In 
past couple of years the successful use of spatial coupling spread from error 
correcting codes to other areas, such as the compressed sensing 
\cite{KrzakalaPRX2012,DonohoJavanmard11}, multiple access communication 
\cite{schlegel2011multiple,takeuchi2011improvement}, group testing 
\cite{zhang2013non}, and others.  The range of applicability is very large 
which also very recently motivated more conceptual studies of spatial
coupling \cite{Urbankechains,YedlaJian12,kudekar2012wave} and the present work belongs 
to that group.

Apart of searching for new applications where spatial coupling can lead to 
improvements, there is a large number of conceptual questions that have not yet 
been answered in a satisfactory manner. For instance what are the conditions--- 
\textit{e.g.}, the size and strength of the seed and the range of coupling--- 
under which a wave propagates and successful inference is reached? What are the 
parameters that control the speed of propagation \cite{aref2013convergence}?  
What are the parameters that lead to smallest possible loss with respect to 
optimality for chains of finite length \cite{kudekar2010threshold}?  What is 
the effect of finite systems size 
\cite{olmos2011scaling,olmos2012finite,olmos2013finite}? The present paper 
contributes to answering these questions and hence towards better understanding 
of the concept of spatial coupling. Above all, such considerations are 
instrumental in practical implementations of the concept in real-world 
applications.

Our approach is to study spatial coupling for the simplest mean field
model with a first order phase transition, that is the Curie-Weiss (C-W) model 
in external magnetic field.  This model contains the most important features of 
more general inference problems whilst remaining analytically simple to treat.  
Indeed, a spatially coupled C-W model was recently introduced by Hassani, 
Macris and Urbanke in \cite{Urbankechains}. Their article contains an excellent 
review of related works and models in the physics literature. Their 
discussion focuses on the equilibrium solutions of the model and showing that 
with spatial coupling one can indeed achieve optimal inference in a tractable 
way. Here, we study the speed of the convergence towards equilibrium--- in 
other words, the speed of the nucleation wave--- and the conditions under which 
the termination conditions leads to a growing nucleus of the equilibrium phase.  
We fully expect that our results will generalize to more complicated and 
practical spatially-coupled problems, such as those described earlier.

Our paper is organized as follows.  In Section \ref{sec:model}, we describe the 
standard C-W model, showing how it corresponds to a general setting of 
inference.  For the most part, this serves as an opportunity to explain and 
introduce some of the terms used in computer science and information theory, 
that are not familiar to the majority of the physics community.  In Section 
\ref{sec:spat_coup}, we consider a chain of such systems that are coupled 
together following \cite{Urbankechains}.  At this stage, Section \ref{sec:VB} 
gives an overview of the two main categories of inference algorithms:
Variational Bayes method and Monte-Carlo.  
In Section \ref{sec:dyn} we derive a continuous differential equation that 
describes the travelling wave and compute its speed.  In Section 
\ref{sec:speed} we then evaluate the speed as a function of various parameters.  
Finally in Section \ref{sec:res} we study the role of the termination 
condition, the range of parameters under which the wave propagates and their 
optimization.  Section \ref{sec:concl} concludes by summarizing our results.

\section{The Curie-Weiss model as an inference problem}
\label{sec:model}

The C-W model in external magnetic field is a textbook example of a mean-field 
model that presents a first order phase transition. The main purpose of this 
section is to briefly introduce the model to non-physics readers and set up the 
analogy with a generic inference problem.  An excellent introduction to the C-W 
model suitable for the present context can also be found in 
\cite{Urbankechains}. 

The C-W model is a system of $N$ Ising spins $s_i\in\{-1,+1\}$, $i=1,\dots,N$, 
that are interacting according to the Hamiltonian of a fully connected Ising 
model
\begin{equation}
	\mathcal{H}_N({\bf s}) = -\frac{J}{N}\sum_{\left\langle i,j\right\rangle} 
	s_i s_j - h \sum_{i=1}^N s_i,
	\label{eq:C-W}
\end{equation}
where the notation $\left\langle\cdot,\cdot\right\rangle$ is used to denote all 
unique pairs, ${\bf s}=\{s_i\ \forall\ i\in\{1,\ldots,N\}\}$,  $J>0$ is the 
ferromagnetic interaction strength and $h\in \mathbb{R}$ is the external 
magnetic field. The Boltzmann probability distribution on spin configurations 
that corresponds to the posterior probability distribution reads 
\begin{equation}
	P_N({\bf s},J,h) = \frac{1}{Z(J,h)}  e^{-\mathcal{H}_N(s)}\, ,
	\label{eq:Boltz}
\end{equation}
where $Z(J,h)$ is the normalization constant, called the partition function in 
physics and the posterior likelihood in inference. The expected value of spin 
$i$ under the measure $P_N({\bf s},J,h)$ is the local magnetization $m_i(J,h)$.  
Moreover for this system, all the local magnetizations are the same, and the 
value $m_i(J,h)=m(J,h)$ is then called the equilibrium magnetization. 

For positive magnetic field $h>0$ the equilibrium magnetization $m(J,h)>0$ is 
also positive, and vice versa. There exists a critical value of the interaction 
strength $J_c=1$ such that: for $J<J_c$ the $\lim_{h\to 0^+} m(J,h) = 
\lim_{h\to 0^-} m(J,h)  = 0$, and for $J>J_c$ we have $\lim_{h\to 0^+} m(J,h) > 
0 > \lim_{h\to 0^-} m(J,h)$. The latter is called a first order phase 
transition, in the `low temperature' regime $J>J_c$ the system keeps non-zero 
magnetization even at zero magnetic field $h$. In mean-field systems, such as 
the C-W model, the first order phase transition is associated with the 
so-called spinodal regime.  There exists a value of the magnetic field
\begin{equation}
	h_{\rm sp}(J)= \sqrt{J(J-1)} -\atanh\left(\sqrt{\frac{J-1}{J}} \right)   \, ,
	\label{eq:hsp}
\end{equation}
with the following properties: if the magnetizations are initialized to
negative values and the magnetic field is of strength  $0 < h < h_{\rm sp}(J)$, 
then both local physical dynamics and local inference algorithms, such as the 
Gibbs sampling or the variational Bayes inference, will stay at negative 
magnetization $m^-(J,h)<0$ forever (or for time exponentially large in the size 
of the system). Hence the spinodal value of the magnetic field $h_{\rm sp}(J)$ 
acts as an algorithmic barrier to equilibration and hence to successful 
inference. For \(h > h_{\rm sp}(J)\) it is, on the other hand, easy to reach the 
equilibrium magnetization $m^+(J,h)$. In the context of error correcting codes 
the phase transition at $h_{\rm sp}(J)$ corresponds to the belief propagation 
threshold \cite{RichardsonUrbanke08}, in the same context the phase transition 
value $h=0$ corresponds to the MAP threshold \cite{RichardsonUrbanke08}.

The first order phase transition and the spinodal region is often explained in 
terms of the free energy, \textit{i.e.}, $f(J,h,m)= - \lim_{N\to 
\infty } \log{Z_N(J,h,m)}$ where $Z(J,h,m)$ is the partition function 
restricted to configurations having magnetization $m=\sum_i s_i/N$.  If the 
external magnetic field $h$ is in the spinodal region $0 < h < h_{\rm sp}(J)$, 
then there are two minima in the free energy, corresponding to 
magnetizations $m^+$ and $m^-$, where the former is the global minimum and the 
latter is only a local minimum.  Furthermore, if $h > h_{\rm sp}(J)$, then only 
one minimum exists which, for consistency, we will still denote $m^+$.

For the purpose of this paper we will always consider ourselves in the `low 
temperature' regime, $J>J_c$. The initial condition for every spin will be 
$s^{t=0}_i=-1$ (unless stated otherwise). The magnetic field $h>0$ will always 
be positive, such that the equilibrium state corresponding to successful 
inference has positive magnetization $m^+(J,h)>0$. For magnetic fields larger 
than the spinodal value $h\ge h_{\rm sp}(J)$, local 
algorithms, such as Monte Carlo sampling or variational Bayes inference (as 
reviewed in Sec.~\ref{sec:dyn}), can reach the equilibrium configuration in a 
number of updates linearly (or log-linearly for random updates) proportional to 
the number of spins. For magnetic fields inside the spinodal region $0< 
h<h_{\rm sp}(J)$ however, such local algorithms will keep negative values of 
magnetization $m^-<0$ for an exponentially long time.  In terms of an analogy 
with inference problems, one can imagine that the magnetic field $h$ is 
proportional to the distance from optimality.  We would therefore like to 
achieve the equilibrium state of positive magnetization in tractable time also 
for $0< h<h_{\rm sp}(J)$. As we will see, and as was shown in 
\cite{Urbankechains}, this is possible with the use of spatial coupling.

\section{Spatially coupled Curie-Weiss model}
\label{sec:spat_coup}

We follow the model definition as set out in \cite{Urbankechains}.  We 
consider a one-dimensional chain of \(2L+1\) C-W systems, where each of the  
C-W system has $n$ spins (referred to as a `block') and is labelled by the 
index $z\in\{-L,\ldots,L\}$.  The result is that a configuration $s$ of the 
full system is now given by the values of $N=n(2L+1)$ spins, each labelled by a 
compound index:
\begin{equation}
	s=\left\{s_{iz}\in\left\{+1,-1\right\}:\ i\in\left\{1,\dots,n\right\}, 
	z\in\left\{-L,\dots,L\right\}\right\}.
	\label{eq:s}
\end{equation}
In a similar way, the uniform external magnetic field $h$ for a single system 
is replaced by an external field profile $h_z$.  As far as the coupling is 
concerned, every spin not only connects to all spins in the same `location' $z$ 
but also all spins within $w$ blocks from $z$. The corresponding Hamiltonian is 
then
\begin{equation}
	\mathcal{H}_{n,L}\left(s\right) =
	-\frac{1}{n}\sum_{\langle iz,jz'\rangle} J_{zz'} s_{iz} s_{jz'} - 
	\sum_{z=-L}^{L} h_z \sum_{i=1}^n s_{iz}.
	\label{eq:H}
\end{equation}
The couplings between spins are 
\begin{equation}
	J_{zz'} = \frac{J}{w} g\left(\frac{\vert z-z'\vert}{w}\right),
	\label{eq:interaction}
\end{equation}
where the function $g$ satisfies the following condition
\begin{equation}
	g\left(\vert x\vert\right) = 0,\ \forall\ \vert x\vert > 1,
	\label{eq:g2}
\end{equation}
and we choose its normalization to be 
\begin{equation}
	\frac{1}{w}\sum_{z=-\infty}^{+\infty} g\left(w^{-1}\vert z\vert\right) = 1.
	\label{eq:g1}
\end{equation}

In the analogy with general inference problems that we discussed in previous 
sections, the average magnetic field $h_{\rm avg} = \sum_z h_z /(2L+1)$ plays 
the role of the cost that we want to minimize while having the low-complexity 
algorithms still able to find the equilibrium state of the system, instead of 
being stuck in the metastable state. (We remind the reader that in general we 
want to consider the initial condition where for every $i$ and $z$ we have 
$s_{iz}=-1$ and $h>0$). However, in order to ensure that the system 
`nucleates', we must increase the field $h_z$ at some point on the chain and 
therefore increase the average values $h_{\rm avg}$.  In this work we choose 
the magnetic field profile in such a way that $h_z = h_\mathrm{seed} > 
h_\mathrm{sp}$ in some small region given by $z\in 
\{0,\ldots,w_\mathrm{seed}\}$--- \textit{i.e.}~$w_\mathrm{seed} = 
\mathrm{meas}\{z: h_z > h_\mathrm{sp}\}$ corresponds to the number of blocks 
covered by the seed.  Everywhere else, $h_z=h \ll h_\mathrm{sp}$, such that the 
average field strength
\begin{equation}
	h_\mathrm{avg} = \frac{w_\mathrm{seed}}{2L+1}(h_\mathrm{seed} - h) + h,
	\label{eq:h_avg}
\end{equation}
is still small.

In most of the theoretical works about spatial coupling, including 
\cite{Urbankechains}, the issue of the seed is circumvented by imposing 
appropriate boundary conditions. In practical cases of inference problems, 
however, the boundary conditions cannot be imposed (see \textit{e.g.}, 
\cite{KrzakalaPRX2012}) and hence the study of necessary properties of the seed 
is essential and so far missing in the literature.

\begin{figure}[t]
	\centering
	\includegraphics[scale = 0.5]{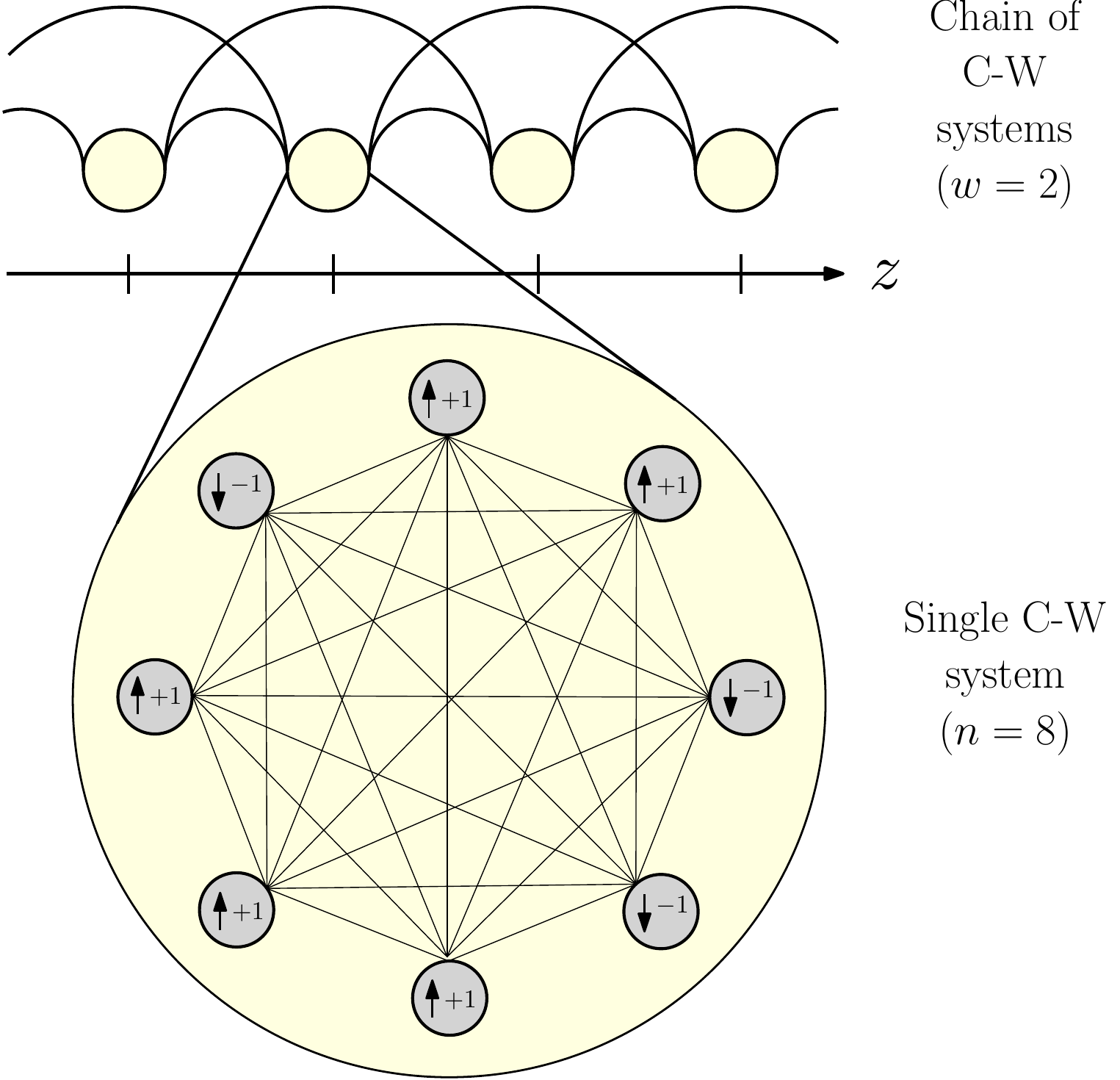}
	\caption{(Color online) A schematic graphical representation of the model.  
		A chain of C-W models interacting within a certain range $w$ ($w=2$ in 
		the figure). In the zoomed part the fully-connected structure of the 
	single C-W model is shown.  (Note that a connection between C-W models 
along the chain indicates connections between all spins contained in both 
models).}
	\label{fig:model}
\end{figure}

\begin{figure}[t]
	\centering
	\includegraphics[scale = 0.95]{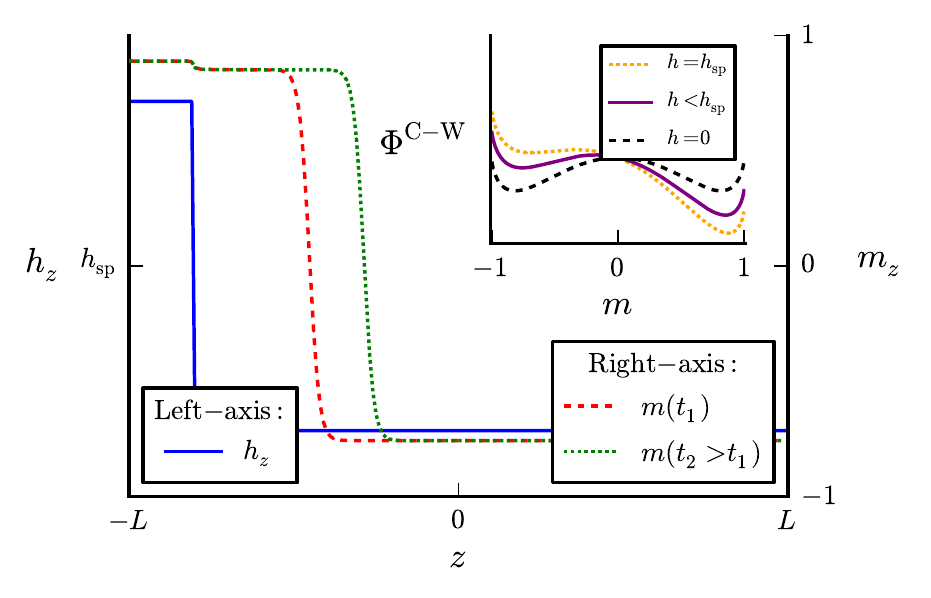}
	\caption{(Color online) A chain of coupled C-W systems.  The external field 
		is a function of position $z$: in the `seed' region $h_z > 
		h_\mathrm{sp}$, whilst in the bulk of the chain $h_z < h_\mathrm{sp}$.  
		Starting every C-W system in the neighborhood of $m^-$, the composite 
		system magnetizes according to a travelling wave that starts at the 
		seed.  Inset: the free energy of a C-W system in a uniform 
		external field.  In the so-called spinodal region, $0<h<h_\mathrm{sp}$, 
		there are two minima $m^-$ and $m^+$, such that $\Phi^\mathrm{C-W}(m^-) 
		>  \Phi^\mathrm{C-W}(m^+)$.  For $h>h_\mathrm{sp}$, $m^+$ is the only minimum 
		of $\Phi^\mathrm{C-W}$.}
	\label{fig:1}
\end{figure}

\section{Variational Bayes vs. Monte-Carlo}
\label{sec:VB}

To solve the above inference problem we need to sample the posterior (Boltzmann) 
probability distribution (\ref{eq:Boltz}) in order to compute its marginal 
probabilities (local magnetizations). The most commonly used methods for 
sampling fall into two classes: variational Bayes and Monte-Carlo (M-C).  
In this paper, we will mostly be concerned with the variational Bayes approach, as it is 
typically faster than M-C.  However, since physicists are familiar with M-C, we 
will fist explain how the two approaches relate. 

In physics, M-C is the `go-to' method for minimizing the free energy 
in a spin system.  Most generally, this involves constructing an algorithm that 
explores the parameter space and iteratively moves towards the state that 
minimizes the energy, subject to entropic constraints.  The process is usually 
designed to be stochastic and Markovian.  Some examples of famous Monte-Carlo Markov-Chain 
(MCMC) methods include the Metropolis-Hastings algorithm or the heat-bath 
algorithm.  The upside of these approaches is that, in the limit of a large 
number of time steps, they are exact.  However, the downside is that 
achieving such accuracy can be computationally expensive.

In terms of the limiting behavior of MCMC algorithms, we can persuade ourselves 
that if we define local magnetization averaged over realizations of the M-C 
dynamics as $m_{iz}^t = \langle  s_{iz}^t \rangle_{\rm realization} $, then for 
the heat-bath algorithm with parallel update, the system evolves according to
\beq
m^{t+1}_{iz}=\tanh \left(   \sum_{z' \neq z} J_{z z'}\sum_{j}
  m^t_{jz'}  +  J_{ z z} \sum_{j\neq i} m^t_{jz} +  h_z\right)\, . \label{eq:MC}
\eeq
Where this expression can also be obtained by analyzing the
M-C heat-bath dynamics in the limit $n=1$ and $w\to \infty$ (instead of
$w$ finite and $n\to \infty$) \cite{de1994glauber}.  If one uses random 
sequential update in the M-C simulation, the behavior of the system in the 
large $N=n(2L+1)$ limit corresponds to the analogous differential equation
\beq
\frac{{\rm d} m_{iz}}{{\rm d}\tau}=\tanh \left(   \sum_{z' \neq z} J_{z 
z'}\sum_{j}
  m_{jz'}  +  J_{ z z} \sum_{j\neq i} m_{jz} +  h_z\right) - m_{iz}\, , 
  \label{eq:MC_cont}
\eeq
in terms of a rescaled time $\tau=t/N$.

In contrast to the above, the variational Bayes algorithm to compute the local marginals 
(magnetizations) uses the assumption that the posterior probability (Boltzmann) 
distribution can be written in a factorized form \beq
         \label{eq:fact}
    P^{\rm MFT}_N({\bf s}, h, J) = \prod_{i,z} \left[ \frac{1+m_{iz}}{2}
    \delta_{s_{iz},1} + \frac{1-m_{iz}}{2} \delta_{s_{iz},-1} \right]
\eeq
When this form is then substituted to the Kullback–Leibler divergence from $ 
P^{\rm MFT}_N({\bf s}, h, J) $ to $P_N({\bf s}, h, J)$ we obtain the free 
energy \beq
\begin{split}
&F=- \sum_{\langle iz ,jz' \rangle}J_{zz'} m_{iz}m_{jz'} -  \sum_{z} h_z 
\sum_{i} m_i \\
&+ \sum_{iz} \left[ \frac{1+m_{iz}}{2} \ln \left( \frac{1+m_{iz}}{2}
  \right) +\frac{1-m_{iz}}{2} \ln \left( \frac{1-m_{iz}}{2} \right)
\right] \\ &= -\log{Z} + D_{\rm KL}( P^{\rm MFT} || P) \, . \label{eq:fe}
\end{split}
\eeq
By imposing a stationarity condition on $F$ with respect to $m_{iz}$, we obtain 
the following fixed-point equation
\beq
 \label{varbayes_fixed}
m_{iz}=\tanh \left(   \sum_{z' \neq z} J_{z z'}\sum_{j}
  m_{jz'}  +  J_{ z z} \sum_{j\neq i} m_{jz} +  h_z\right)\, .
\eeq
Solutions to this equation can be found, for instance, by iterating 
(\ref{eq:MC}).  So we see that for the C-W model, the evolution of the heat-bath 
MCMC is closely related to the variational Bayes inference. Indeed, it is also 
worth noting that, for the C-W model, the variational Bayes approach provides asymptotically 
exact values of the marginals (local magnetizations). For general inference 
problems no such algorithm exists, whilst for the two main motivational 
examples of this paper--- LDPC codes and compressed sensing--- the Bethe 
approximation (and a related belief propagation algorithm) is asymptotically 
exact.

\begin{figure}[t!]
	\centering
	\includegraphics[scale=0.7]{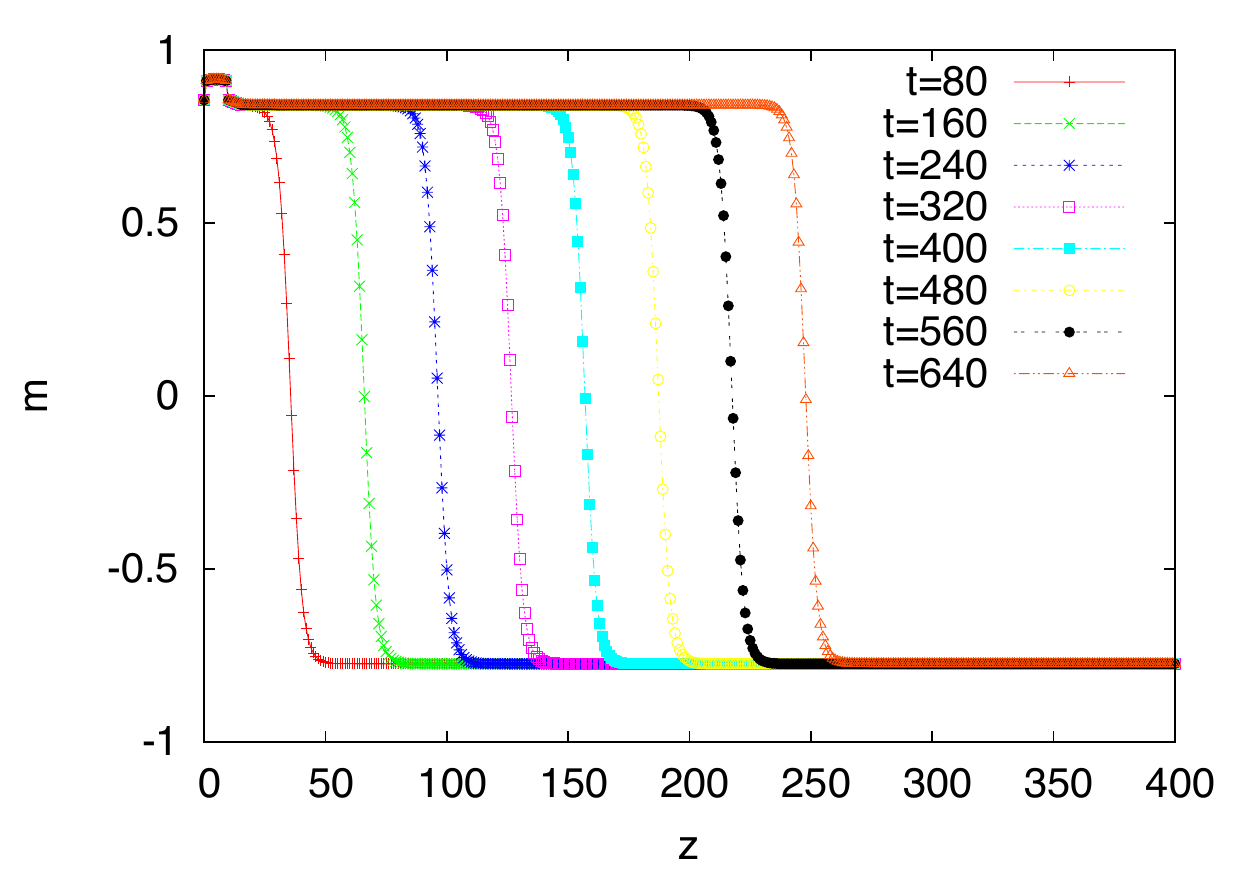}
	\caption{(Color online) The propagating wave obtained by iteration of 
		Eq.~(\ref{eq:MC}) for the following parameters: $J=1.4$, $L=400$, 
		$n=40$, $h=0.05$, $w=5$, $w_{\mathrm{seed}}=10$ and 
		$h_{\mathrm{seed}}=0.3$.  
}
	\label{fig:2}
\end{figure}

\begin{figure}[t!]
	\centering
	\includegraphics[scale = 0.7]{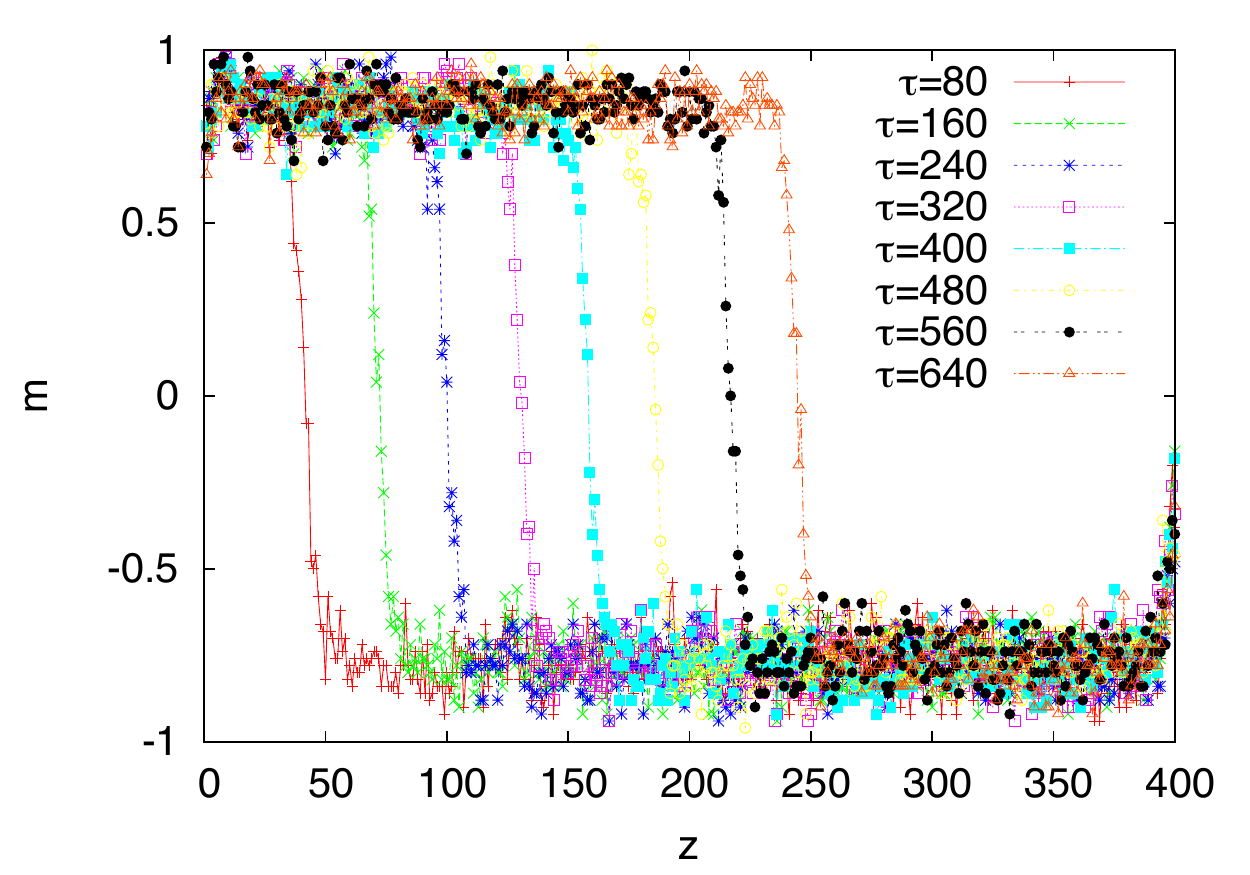}
	\caption{(Color online) The propagating wave obtained by
		Monte-Carlo heat-bath simulation for the following parameters: $J=1.4$, 
		$L=400$, $n=100$, $h=0.05$, $w=5$, $w_{\mathrm{seed}}=10$ and 
		$h_{\mathrm{seed}}=0.3$.  Simulations were performed using random 
	sequential update and therefore time must be rescaled by
        $\tau=t/N$.}
	\label{fig:MC_waves}
\end{figure}

As anticipated, iterating (\ref{eq:MC}) (with appropriate parameter choices) 
results in a travelling wave of magnetization emanating from the seeded region, 
and eventually saturating the whole chain to the global optimum, $m^+$.  As an 
example, in Fig.~\ref{fig:2} we show the propagating wave for the following 
parameter values: $\,J=1.4\,,\,h=0.05\,,\,w=5\,,\, n=40\,,\,L=400$, where for 
this value of $J$, the field is below the spinodal point $h_{\rm sp}\approx 
0.1518$.  We initialize the whole system with local negative magnetization and 
on the first $w_{\mathrm{seed}}=10$ blocks we impose a seeding field 
$h_{\mathrm{seed}}=0.3$, which is above $h_{\rm sp}$. As can be seen in the 
figure, we observe the propagation of a magnetization wave. 

So far, we have only discussed equations in terms of local magnetizations
of a single spin variable. However, it is straightforward to see that,
since we have no disorder, the local magnetization does not fluctuate from 
site-to-site within one block. We can therefore derive the equivalent of the 
`state' or 'density' evolution for the above algorithm by simply summing over $i$ on both 
sides of the equation, dividing by $n$, and then letting $n \to \infty$.  In 
this way we obtain
\beq
\label{stateevo}
m^{t+1}_z= \tanh \left(  \sum_{ z' } D_{z z'} m^t_{z'} + J  m^t_z +  h_z  
\right) ,
\eeq
where
\beq
D_{z z'}=J_{z z'}-\d_{z,z'} J.
\eeq
Here, it is helpful to make a few remarks: firstly, in the C-W model, the 
finite size fluctuations--- neglected in (\ref{stateevo})--- are just a small 
(deterministic) correction to the fixed point values of the local 
magnetizations.  This is different in more realistic problems like LDPC or 
compressed sensing where the disorder (either in the factor graph or in the 
measurement matrix) induces finite size effects that are not captured by the 
C-W model.  Secondly, the free energy (\ref{eq:fe}) can be minimized by 
iterative schemes other than (\ref{eq:MC}), and also other MCMC algorithms 
lead to different evolution equations (\ref{eq:MC}) with the same fixed point.  
For example, the Metropolis-Hastings MCMC with random sequential update is equivalent 
to 
\beq
   \frac{{\rm d} m_{z}}{{\rm d}\tau} =2 e^{- \tilde{h}_{z}} \cosh\left(\tilde{h}_{z}\right)  \left[ \tanh 
	   \left(  \tilde{h}_z\right) -
m_{z} \right],
\label{eq:met}
\eeq
with
\beq
\tilde{h}_z=\sum_{ z' } D_{z z'} m_{z'} + J  m_z +  h_z
\eeq  
in the large system-size limit.  
Whilst taking the steepest descent of the free energy (\ref{eq:fe}) leads
to
\beq
m_{z}^{t+1}= \sum_{ z' } D_{z z'} m^t_{z'} + J  m^t_z +  h_z - \atanh 
(m^t_{iz})+m^t_{iz}.
\label{eq:steep}
\eeq
In this paper, we pay specific attention to the most common update 
(\ref{eq:MC}), but we stress that the same kind of analysis can be carried out 
for (\ref{eq:met}-\ref{eq:steep}) and others.

\section{Continuous limit}
\label{sec:dyn}

Useful information about the behavior of the spatially coupled system can be 
understood if, following \cite{Urbankechains},  we take the double limit $L\gg 
w\gg 1$, such that $L,w\to +\infty$ whilst $w/L\to 0$.  
In addition to this, 
introduce a space variable $x = z/w$ 
such that the limiting magnetization profile
\begin{equation}
	\lim_{L,w \to +\infty :\ w / L\to 0} m_{wx}(\tau) \equiv m(x, 
	\tau)\in[-1,+1],
	\label{eq:m_of_tau}
\end{equation}
is now a function of the continuous variables $x, \tau\in [-\infty,+\infty]$.  
In this limit, the sum over the kernel $D_{zz'}$ becomes
\begin{equation}
	\lim_{\substack{w,L\to +\infty\\ w / L\to 0}}\sum_{z'=-L}^L 
	D_{zz'}m_{z'}(\tau) = \int_{-\infty}^{+\infty} \mathrm{d}x' D\left(x'\right) 
	m(x-x',\tau),
	\label{eq:kernel_limit}
\end{equation}
where $
D\left(x\right) = J\left[ g\left(\vert x\vert\right) - 
	\delta\left(x\right)\right]
$, and the right-hand side has been re-written using change of variable 
$x'\mapsto x-x'$.  The state evolution equation (\ref{stateevo}) can then be
written in the continuous limit as
\begin{equation}
	\frac{\mathrm{d} m}{\mathrm{d} \tau} = \tanh\left(\left[ D*m\right] + 
	 J m +  h\right) - m,
		\label{eq:BP_cont}
\end{equation}
where the shorthand $[\cdot *\cdot]$ represents convolution and the explicit 
dependencies of $m(x,\tau)$ and $h(x)$ have been dropped.
Notice that when $\vert x\vert \gg 1$, $D(x)=0$ and therefore in the convolution we can perform a gradient expansion.  Due to symmetry, the lowest order term in the expansion with 
a non-zero contribution is quadratic.  That is, the right-hand side of 
(\ref{eq:kernel_limit}) is approximated by
\begin{equation}
	\sim\left(\frac{J}{2} \int_{-\infty}^{+\infty} \mathrm{d}x' g\left(\vert 
	x'\vert\right) x'^2 \right) \frac{\partial^2 m}{\partial x^2} + 
	\mathcal{O}(x^3).
	\label{eq:taylor}
\end{equation}
Substituting into (\ref{eq:BP_cont}) and inverting the hyperbolic tangent then gives the result
\begin{equation}
	\omega \frac{\partial^2 m}{\partial x^2}=-  J m -  h + \atanh\left( m+\frac{\partial m}{\partial \tau}\right)
		\label{eq:R-D}
\end{equation}
where the pre-factor from (\ref{eq:taylor}) has been absorbed in the definition
\begin{equation}
	\omega = \frac{J}{2} \int_{-\infty}^{+\infty} \mathrm{d}x' 
	g\left(\vert x'\vert\right) x'^2.
	\label{eq:omega}
\end{equation}

Working in this limit, it is now possible to look for traveling-wave 
solutions--- that is, those of the form $m(x,\tau) = m(x - v\tau)$.  This leads 
to an ordinary differential equation in terms of variable $y=x-v\tau$:
\begin{equation}
	\omega\frac{\mathrm{d}^2 m}{\mathrm{d}y^2} =
	- J m -  h + \atanh\left(m-v\frac{\mathrm{d} m}{\mathrm{d} 
y}\right).
\label{eq:trav_wave}
\end{equation}
If the bulk field $h$ is very small, the potential barrier between 
$m^-$ and $m^+$ is large and therefore $v$ is expected to be small too.  The 
resulting expansion for $\vert v\vert\ll1$ gives
\begin{equation}
	\omega\frac{\mathrm{d}^2 m}{\mathrm{d}y^2} \simeq \frac{\partial 
		\Phi^\mathrm{C-W} (m)}{\partial m} - \frac{v}{1-m^2}\frac{\mathrm{d} 
	m}{\mathrm{d} y}.
\label{eq:small_v}
\end{equation}
where $\Phi^\mathrm{C-W}$ is the free energy associated to the single
C-W model, \textit{i.e.}, Eq.~(\ref{eq:fe}) for $L=0$. Multiplying both sides 
of this equation by $\mathrm{d} m(y) / \mathrm{d} y$, integrating over 
$y\in\mathbb{R}$, and applying the boundary conditions $m(-\infty) = m^+$, 
$m(+\infty) = m^-$, and $\mathrm{d} m(y) / \mathrm{d} y \vert_{y=-\infty} = 
\mathrm{d} m(y) / \mathrm{d} y \vert_{y=+\infty} = 0$ implies that
\begin{equation}
	v \simeq \Delta \Phi^\mathrm{C-W} / \gamma,
	\label{eq:v_main}
\end{equation}
where the numerator $\Delta\Phi^\mathrm{C-W}$ is just the difference between 
the two minima of the C-W potential
\begin{equation}
	\Delta\Phi^\mathrm{C-W} = \Phi^\mathrm{C-W}(m^-) - \Phi^\mathrm{C-W}(m^+),
	\label{eq:DeltaPhi_main}
\end{equation}
and the denominator $\gamma$ is given by
\begin{equation}
	\gamma = \int_{-\infty}^{+\infty} 
	\frac{\mathrm{d}y}{1-m^2(y)}\left[\frac{\mathrm{d} m(y)}{\mathrm{d} 
	y}\right]^2.
	\label{eq:gamma_main}
\end{equation}
An approximation for $\gamma$ that is consistent with a small $v$ assumption 
can be found by setting $v=0$ in (\ref{eq:small_v}).  Once again multiplying 
both sides by $\mathrm{d} m(y) / \mathrm{d} y$, and this time integrating $y$ 
from $0$ to $Y$, the result is that
\begin{equation}
	\frac{\omega}{2}\left[\frac{\mathrm{d} m(y)}{\mathrm{d} y}\right]^2 \simeq 
	\Phi^\mathrm{C-W}(m(y)) - \Phi^\mathrm{C-W}(m^-),
	\label{eq:v=0}
\end{equation}
where the aforementioned boundary conditions have been applied and $Y$ has been 
relabeled $y$.  Substituting back into (\ref{eq:gamma_main}) gives
\begin{equation}
	\gamma\simeq\int_{m^-}^{m^+}\frac{\mathrm{d}m}{1-m^2}
	\left\{
		\frac{2\left[ 
			\Phi^\mathrm{C-W}(m) - 
			\Phi^\mathrm{C-W}(m^+)
		\right]}{\omega}
	\right\}^{1/2}.
\label{eq:gamma_small_v}
\end{equation}
From (\ref{eq:gamma_small_v}) and (\ref{eq:v_main}) we can easily obtain an 
analytic formula for $v$ as a linear expansion in small $h$ \beq
v=\frac{ (m^+_0-m^-_0)}{\int_{m^-_0}^{m^+_0}\frac{\mathrm{d}m}{1-m^2}
	\left\{
		\frac{2\left[ 
			\Phi_0^\mathrm{C-W}(m) - 
			\Phi_0^\mathrm{C-W}(m^+_0)
		\right]}{\omega}
	\right\}^{1/2}} \, h
	\label{linear}
\eeq
where the suffix `0' refers to quantities at zero bulk field.

Note that this last Eq.~(\ref{linear}) and the whole derivation of this Section 
is closely related to the description of flat moving interface in statistical 
physics of nucleation, see \textit{e.g.},~\cite{godreche1991solids}. We also 
mention here that the steepest descent update (\ref{eq:steep}) would lead to a 
different continuous equation, namely
\begin{equation}
	\omega \frac{\partial^2 m}{\partial x^2}=-  J m -  h + \atanh\left( 
	m\right) +\frac{\partial m}{\partial \tau},
		\label{eq:steep_cont}
\end{equation}
that has the form of a bistable reaction-diffusion equation.

\section{Front velocity as a function of the bulk field}
\label{sec:speed}

We have already seen that one can calculate an approximation to the velocity of 
propagation of the wave--- a proxy for the rate of convergence in inference--- 
by taking an appropriate continuous limit and assuming $h$ to be small.  
However, it is helpful to know how good this approximation is, and indeed if 
there are any other factors that affect the relationship between $v$ and the 
bulk field~$h$.  (Recall that, in the analogy between the C-W model and generic 
inference problems, the value of $h$ represents a distance from the optimal 
setup).
  
\subsection{Large $w$, flat interaction.}  
  
\begin{figure}[h]
	\centering
	\includegraphics[scale = 0.7]{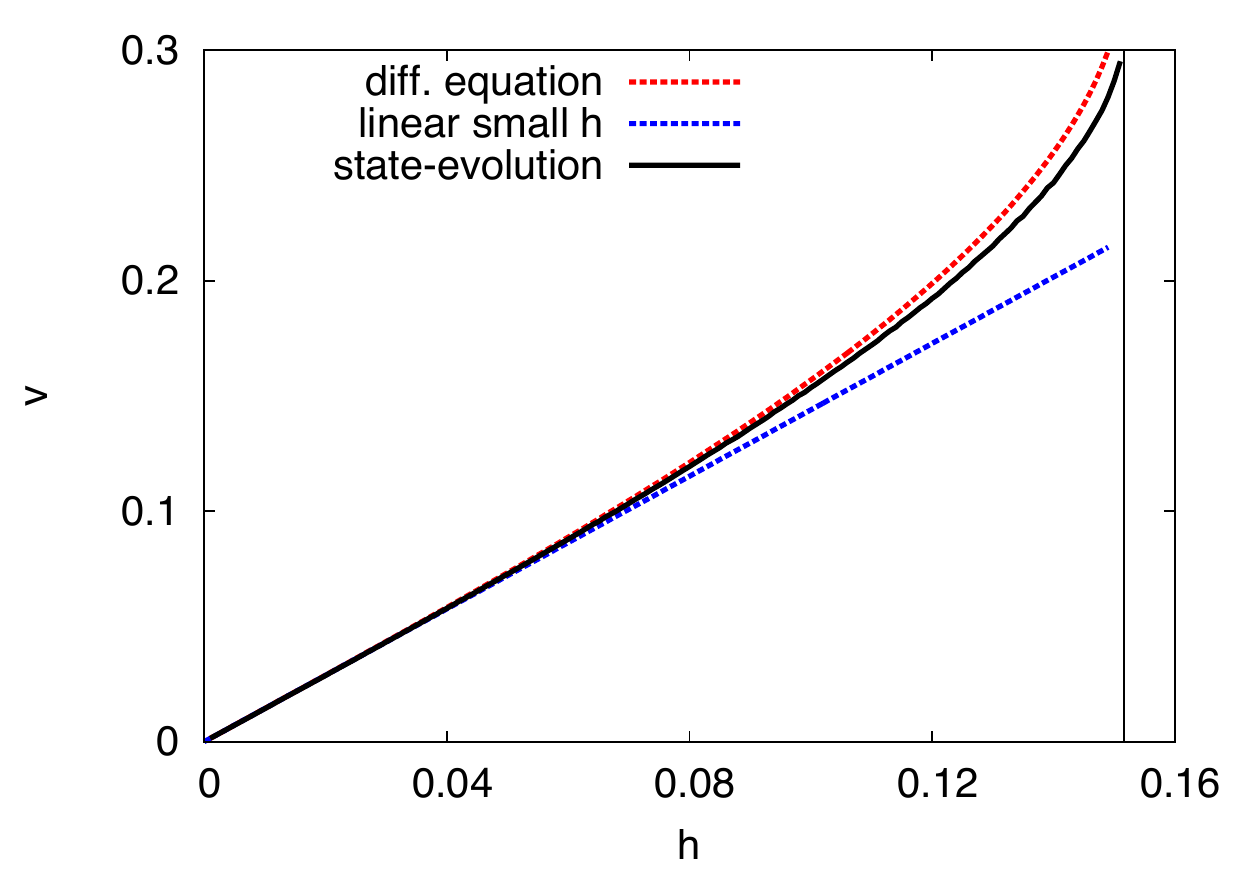}
	\caption{(Color online) The normalized velocity $v=\dot{z}/w$ as a function 
	of the bulk field $h$ for $J=1.4$ computed in three different ways. The 
	solid line is the `experimental' speed measured in the iteration of 
	Eq.~(\ref{stateevo}) ($L=1000$, $w=30$), the dotted curved line is the 
	speed determined through integration of Eq.~(\ref{eq:trav_wave}) and the 
	dotted straight line is given by the linearized analytic formula 
	(\ref{linear}).}
	\label{fig:speedVSh}
\end{figure}
  
In Fig.~\ref{fig:speedVSh}, for the simplest case in which the function $g$ is 
a constant, we compare three different curves:
\begin{enumerate}
	\item The velocity in the limit $n\gg L\gg w\gg 1$ as given by the ordinary 
		differential equation (\ref{eq:trav_wave}). In order to obtain a
		(numerical) value for the propagation speed $v$ we integrate 
		(\ref{eq:trav_wave}) between $-T$ and $T$ with $T\gg 1$,
		with an initialization $m(-T)=m^+$ and $m'(-T)\approx 0$.
	
	\item The velocity in the limit $n\gg L\gg w\gg 1$ as calculated from the 
		analytic linearized formula given in Eq.~(\ref{linear}).
	
	\item The velocity computed from numerical iteration of the asymptotic  
		update (\ref{stateevo}) at finite length $L=1000$, and finite 
		interaction range $w=30$.  In this case, we only use parameter values 
		that lead to wave propagation (propagation/non-propagation conditions 
		will be discussed in the next Section).  Note that, given that the wave 
		propagates, its speed (in the bulk) is independent
                (within the numerical precision) of the seed.
\end{enumerate}
The main result here is that, as expected, the asymptotic update equation in 
the continuous limit gives a good estimate for values of the field not too 
close to the spinodal point.  Indeed, both results are bounded from below by 
the analytical form (\ref{linear}), valid for sufficiently small $h$.


\subsection{Large $w$, different shapes of the interaction.}

\begin{figure}[h]
	\centering
	\includegraphics[scale = 0.7]{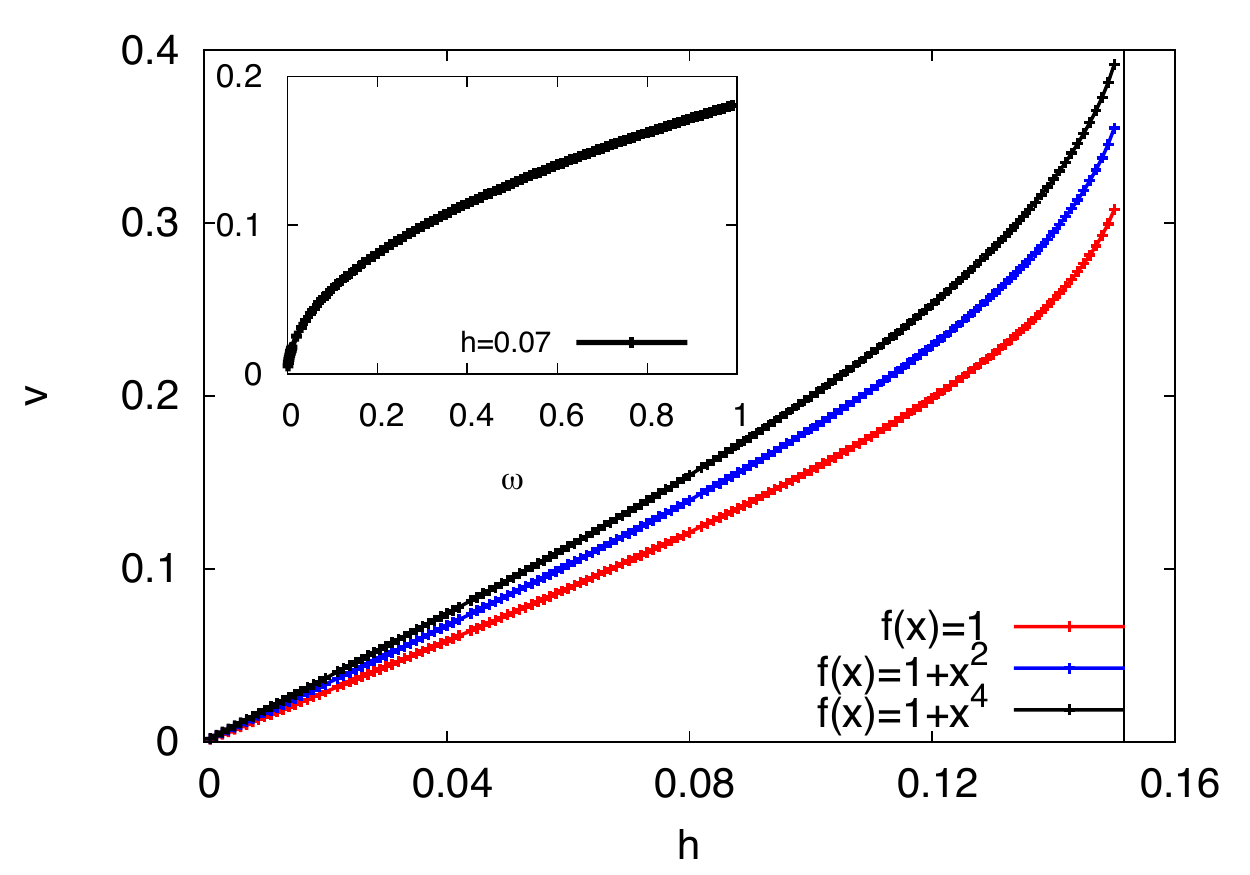}
	\caption{(Color online) [Main frame] The normalized velocity $v=\dot{z}/w$ as a function 
	of the bulk field $h$ with $J=1.4$ for the three model variants described 
	in the main text. From bottom to top: tophat interaction, quadratic interaction 
	and quartic interaction.  [Inset] The speed computed by integration of 
	Eq.~(\ref{eq:R-D}) for fixed bulk field $h=0.07$ and varying $\omega$. }
	\label{fig:speedVSg}
\end{figure}

So far, our analysis has only considered one type of interaction, that gives 
equal weights to all neighbors in a certain range--- \textit{i.e.}, $g$ is a 
`tophat' function.  In fact, the velocity of the front can be increased by 
changing the `shape' of this interaction such that the coupling is strongest 
between blocks that are furthest away (but still bounded by the limits of the 
interaction range $w$).  In order to show this, consider the generalized form
\begin{equation}
	g(x)=\left\{
		\begin{array}{cl}
		0\ &\mathrm{if}\ |x|>1 \\
		f(x)\left[\int_{-1}^1 f(x){\rm d}x\right]^{-1} &\mathrm{if}\ |x|\leq 1
	\end{array}
	\right.
	,
	\label{eq:g_general}
\end{equation}
where now $f$ can be any symmetric function.  To demonstrate the effect of $f$, 
Fig.~\ref{fig:speedVSg} shows the velocity computed by integrating 
(\ref{eq:trav_wave}) for three different forms: $f\sim\mathrm{constant}$ 
(tophat), $f\sim 1+x^2$ (parabolic) and $f\sim 1+x^4$ (quartic).  In the inset 
of Fig.~\ref{fig:speedVSg}, we also show the full dependence of the speed on 
the constant $\omega$--- Eq.~(\ref{eq:omega})--- for a given value of the bulk 
field.  Indeed, from Eq.~(\ref{linear}), we see explicitly that, in the regime 
in which the linearization is a consistent approximation, $v\sim\sqrt{\omega}$. 

It is important to stress at this point that changing $\omega$ is the only tool 
that we have to truly maximize the speed (in terms of `physical time').  To see 
this, suppose that we are at fixed chain length $L$, block spins $n$, bulk 
field $h$, and our unit of time is the single iteration over the whole system.  
We have already stated that, once propagation is achieved, the speed is 
proportional to $w$ and independent of the shape and extent of the seed. On the 
other hand, the single iteration takes a time which is linear in $n$, $L$ and 
$w$, therefore changing the interaction range has no effect on the speed in 
terms of real time and the only effect on the physical speed is given by the 
{\it shape} of the interaction, which in turn determines the value of $\omega$.

\subsection{Finite $w$ effects.}

Until this point we have only considered parameter values that are close to the 
continuous limit--- \textit{i.e.}, $L\gg w \gg 1$.  However, as will 
become clear in the following Section, the case of small $w$ is very relevant 
for practical implementations of spatial coupling.  Indeed, in previous work 
\cite{Urbankechains}, it has been shown that small oscillations in the Van der 
Waals curve mean that propagation can only be formally proved for $w\gg1$.  
However, simulations indicate that the reality of the situation is much more 
positive, and small values of $w$ might be practical in many situations.

\begin{figure}[h]
	\centering
	\includegraphics[scale = 0.7]{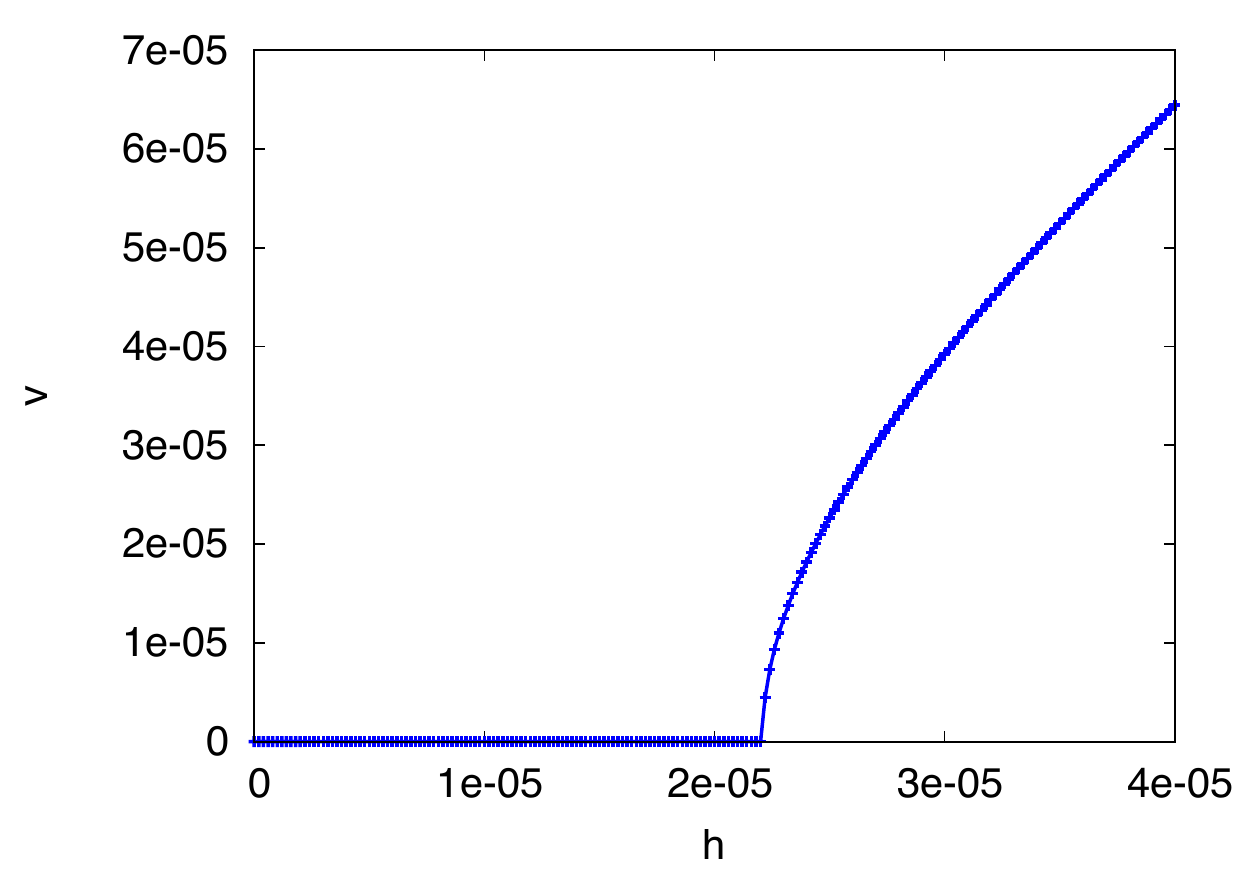}
	\caption{(Color online) The normalized velocity as a function
          of the bulk field $h$ for spatially coupled
          system with interaction range $w=1$. For bulk fields smaller
        than the critical value $h<A_1$ the wave does not propagate.}
	\label{fig:wiggles}
\end{figure}

The authors of \cite{Urbankechains} consider the global average magnetization 
along the chain, namely
\beq
m=\frac{1}{2L+1} \sum_{z=-L}^{L} m_{z}
\eeq 
and show that the propagation of the wave (regardless of the seeding scheme) is 
subject to the constraint that the free energy has non-positive derivative with 
respect to $m$ between the two uniform states $m^-_0$ and $m^+_0$--- the 
negative and positive equilibrium states respectively of the uncoupled C-W 
model at $h=0$.  With uniform interaction, $J>1$, and small magnetic field, the 
derivative of the free-energy reads \cite{Urbankechains}
\beq
C(w,J) \, e^{-\frac{\pi^2 w}{J m_0^+}} \sin \left( 2\pi \frac{m}{m_0^+} L 
\right) -h
\eeq
where $C(w,J)$ is a pre-factor that can be computed. This means that we need 
(regardless of the seeding conditions)
\beq
h \geq A_w \equiv C(w,J) \, e^{-\frac{\pi^2 w}{J m^0_+}}
\eeq
for the derivative to be negative everywhere and propagation to be possible. In 
fact, the authors of \cite{Urbankechains} evaluated the amplitude $A_w$ of the 
oscillation for $J=1.4$ in their Table 2, and for $w=1$ they find 
$A_1=2.2\times 10^{-5}$.  In Fig.~\ref{fig:wiggles} we show the speed of the 
wave for $J=1.4$ at very weak bulk field with ideal seeding conditions 
(\textit{i.e.}, we fix half of the chain to positive magnetization).  We see 
that at small bulk field $h<A_w$ the velocity is zero, is agreement with the 
theory of \cite{Urbankechains}.  However, more broadly, it is clear that this 
effect is observable only at extremely small magnetic fields, even for $w=1$.  
Furthermore, the effect decreases exponentially with the value of $w$.  We 
therefore conclude that these oscillations are not likely to cause problems in 
practical implementations of spatial coupling.

\section{The role of the termination condition}
\label{sec:res}
As discussed in the Introduction, the majority of existing studies regarding 
wave propagation in spatially coupled systems consider initial conditions 
analogous to  $m(-\infty) = m^+$, $m(+\infty) = m^-$, and $\mathrm{d} m(y) / 
\mathrm{d} y \vert_{y=-\infty} = \mathrm{d} m(y) / \mathrm{d} y 
\vert_{y=+\infty} = 0$.  In real inference problems however, there is no way of 
fixing such an initial condition and instead the propagation of the wave has to 
be enforced by a proper termination condition--- \textit{i.e.}, a `seed', as 
introduced in Sec.~\ref{sec:spat_coup}. In this Section we analyze the 
properties of the seed under which a travelling wave starts to propagate in the 
system, and we discuss how to optimize its cost (corresponding to the
rate loss in coding).

\subsection{Conditions for propagation}
\begin{figure}[]
	\centering
	\includegraphics[scale = 0.7]{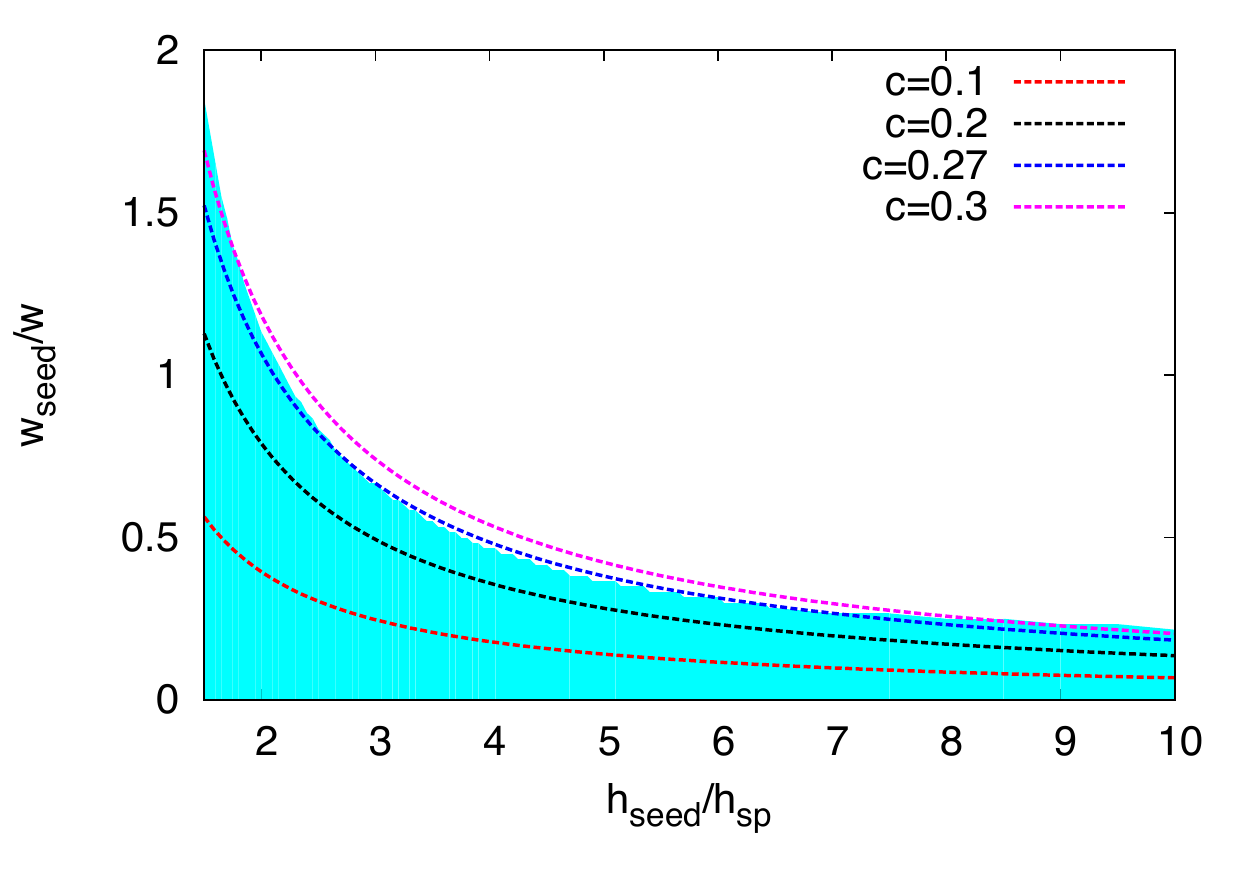}
	\caption{(Color online) The threshold between propagation and
		  non-propagation for $J=1.4$, $w=60$ and $h=0.05$ given by the state 
		  evolution, Eq.~(\ref{stateevo}). Cyan (light grey) represents region 
		  of parameters where the wave does not propagate and white represents 
		  a region where it does propagate.  The dotted lines superimposed to 
		  the shading are cost-density isolines, defined by 
		  Eq.~(\ref{eq:cost}), for different values of $c$.}
	\label{fig:phase}
\end{figure}

The travelling wave will start to propagate in the spatially coupled
system only if the size of the seed ($w_{\rm seed}$) and the strength of the 
seed ($h_{\rm seed}$) are both large enough. Indeed, the 
propagation/non-propagation boundary can be plotted as a function of 
(normalized) variables $h_{\mathrm{seed}}/h_{\rm sp}$ and $w_{\mathrm{seed}}/w$ 
(see Fig.~\ref{fig:phase}).

\begin{figure}[t!]
	\centering
	\includegraphics[scale = 0.7]{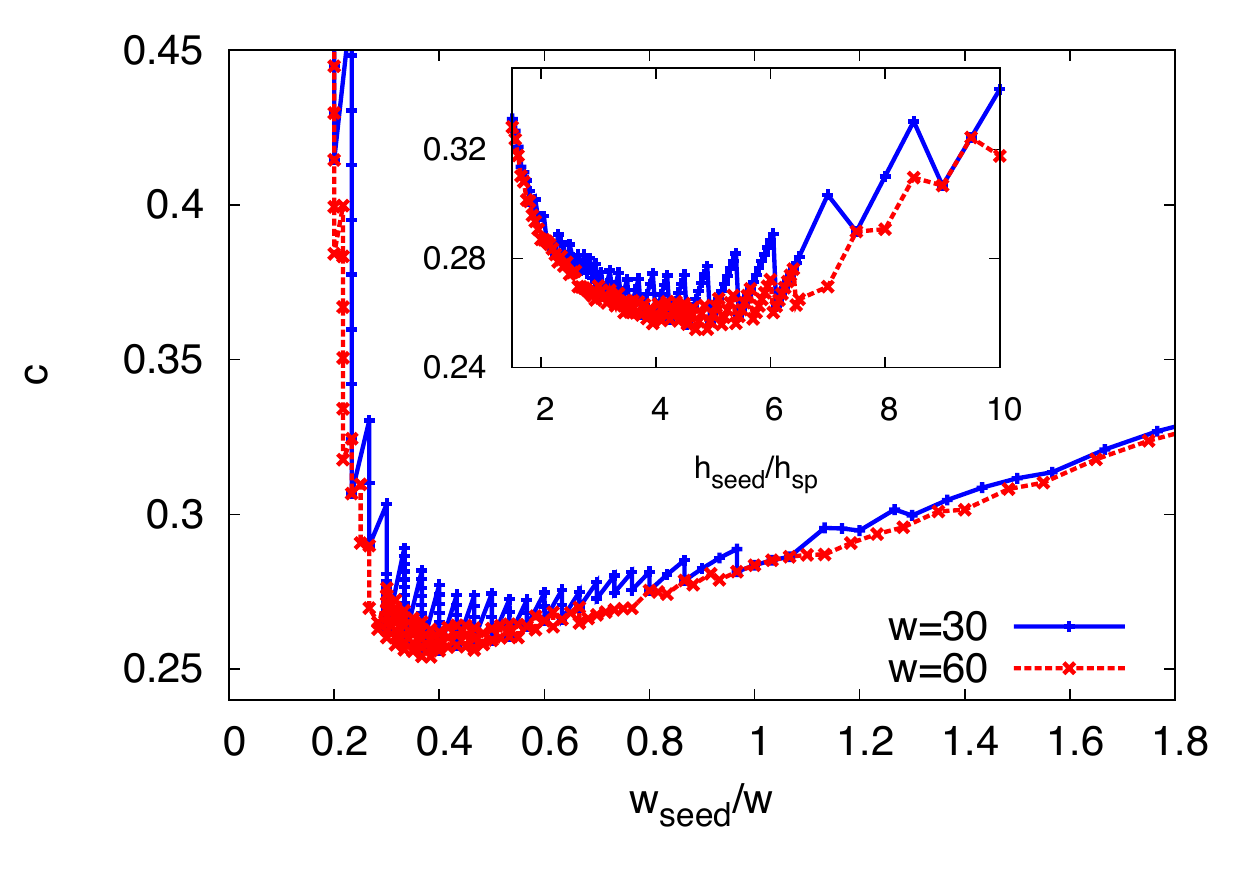}
	\caption{[Main Frame] (Color online) The cost-density along the threshold 
		line in Fig.~\ref{fig:phase} as a function of the ratio 
		$w_{\mathrm{seed}}/w$, for $w=30$ and $w=60$. [Inset] (Color online) 
		The cost-density along the threshold line in Fig.~\ref{fig:phase} as a 
		function of the seeding field, for $w=30$ and $w=60$. In both cases, the 
	plots tend to smooth curve, independent of $w$, for $w\gg 1$.}
	\label{fig:cost2}
\end{figure}

When designing the spatially coupled system, our main objective 
is to obtain propagation of the magnetization wave while keeping 
the average magnetic field on the chain as close as possible to the bulk field 
$h$--- \textit{i.e.}, the goal is to minimize the cost
\beq
\begin{split}
\CC &=h_{\rm av}-h=w_{\mathrm{seed}}
\frac{h_{\mathrm{seed}}-h}{2L+1} = \\
&= \frac{w}{2L+1}
c\left(\frac{w_{\mathrm{seed}}}{w}, \frac{h_{\mathrm{seed}}}{h_{\rm
		sp}} \right),
	\end{split}
\label{eq:cost}
\eeq
which is analogous to the rate loss in error-correcting codes, for example.  
Here, $c$ is a `cost density', and is a function of rescaled variables.  This 
means that for a given choice of the chain-length and interaction range, the 
cost-density isolines can be superimposed in rescaled variables.  As can be 
seen in Fig.~\ref{fig:phase}, all but one of the isolines cross the threshold 
line twice. Given $w$ and $L$ fixed then, the best choice of parameters is 
defined as the point where the unique cost isoline and the threshold line meet 
tangentially.  In Fig.~\ref{fig:cost2} we show how $c$ varies on the threshold 
line as parametrized by $h_{\mathrm{seed}}/h_{\rm sp}$ and 
$w_{\mathrm{seed}}/w$.  From this it is clear that there is indeed a choice 
that corresponds to minimal cost.

Fig.~\ref{fig:cost2} shows explicitly that the cost density curve along the 
threshold line tends towards being independent of $w$ at large values $w$.  In 
other words, for $w \gg 1$, the cost of achieving propagation depends only on 
the ratio $w_{\mathrm{seed}}/w$ and not on $w$ itself.  This kind of analysis 
gives more insight in the choice of parameters that achieve propagation 
minimizing the `field loss' (or rate loss in the case of coding theory), and 
can function as a guide in practical code implementations when $w$ is large.


\subsection{Termination cost at small $w$}

Analysis of the previous section is not valid for small $w$, when the effect 
of the intrinsic discreteness of the system is strong. At the same
time the total cost, eq.~(\ref{eq:cost}),  is linearly dependent
on the range $w$, consequently it seems important to evaluate the cost at
small values of $w$. In this Section we hence consider the typical case of a chain 
of length $L=500$ with $J=1.4$ and small values of the interaction
range $w$. 
Making use of the state evolution (\ref{stateevo}) we study what is the minimal 
cost for propagation as a function of the bulk field for different values of $w$.

\begin{figure}[t]
	\centering
	\includegraphics[scale = 0.7]{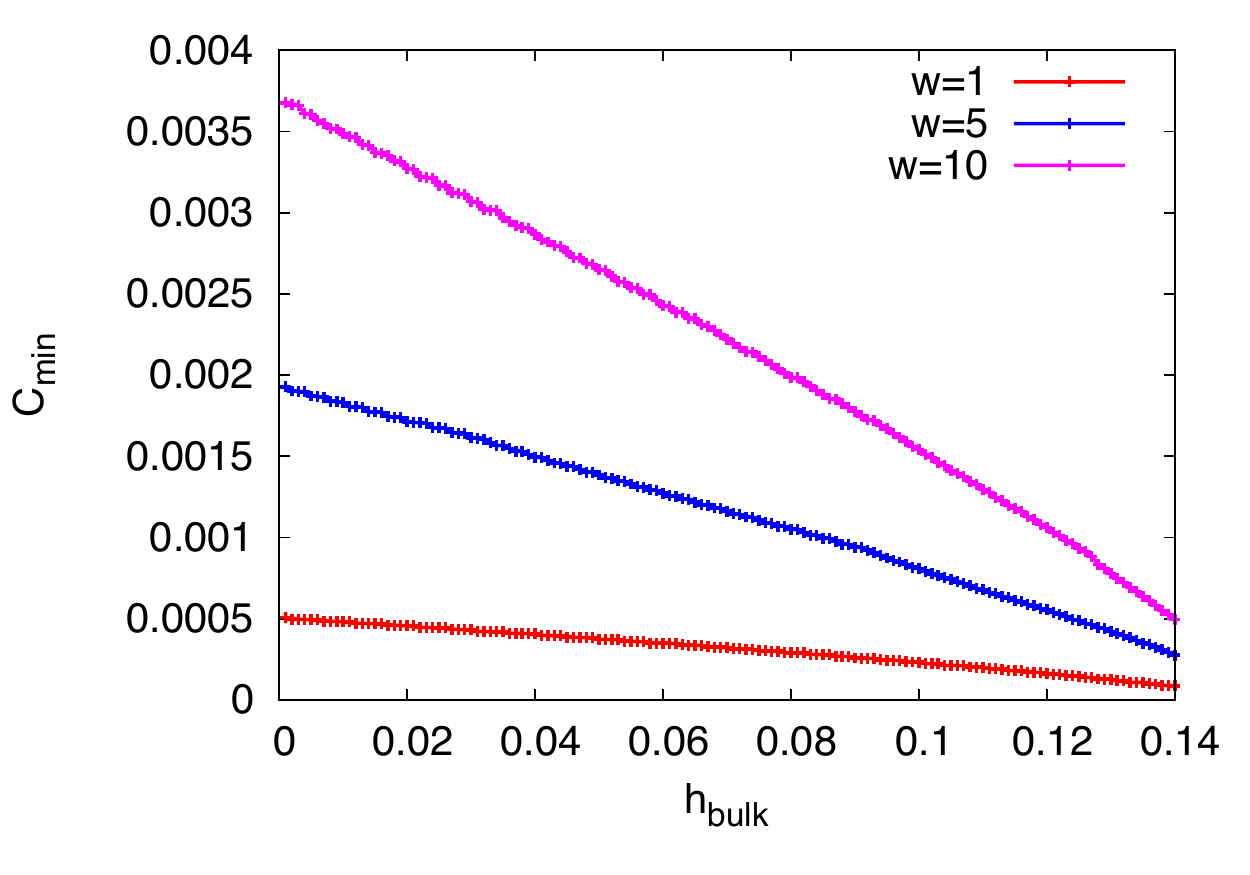}
	\caption{(Color online) The minimal cost as a function of the bulk field for $L=500$, $J=1.4$ and different values of the interaction range. In particular, from bottom to top, $w=1$, $w=5$ and $w=10$.}
	\label{fig:mincost}
\end{figure}

\begin{figure}[t]
	\centering
	\includegraphics[scale = 0.5]{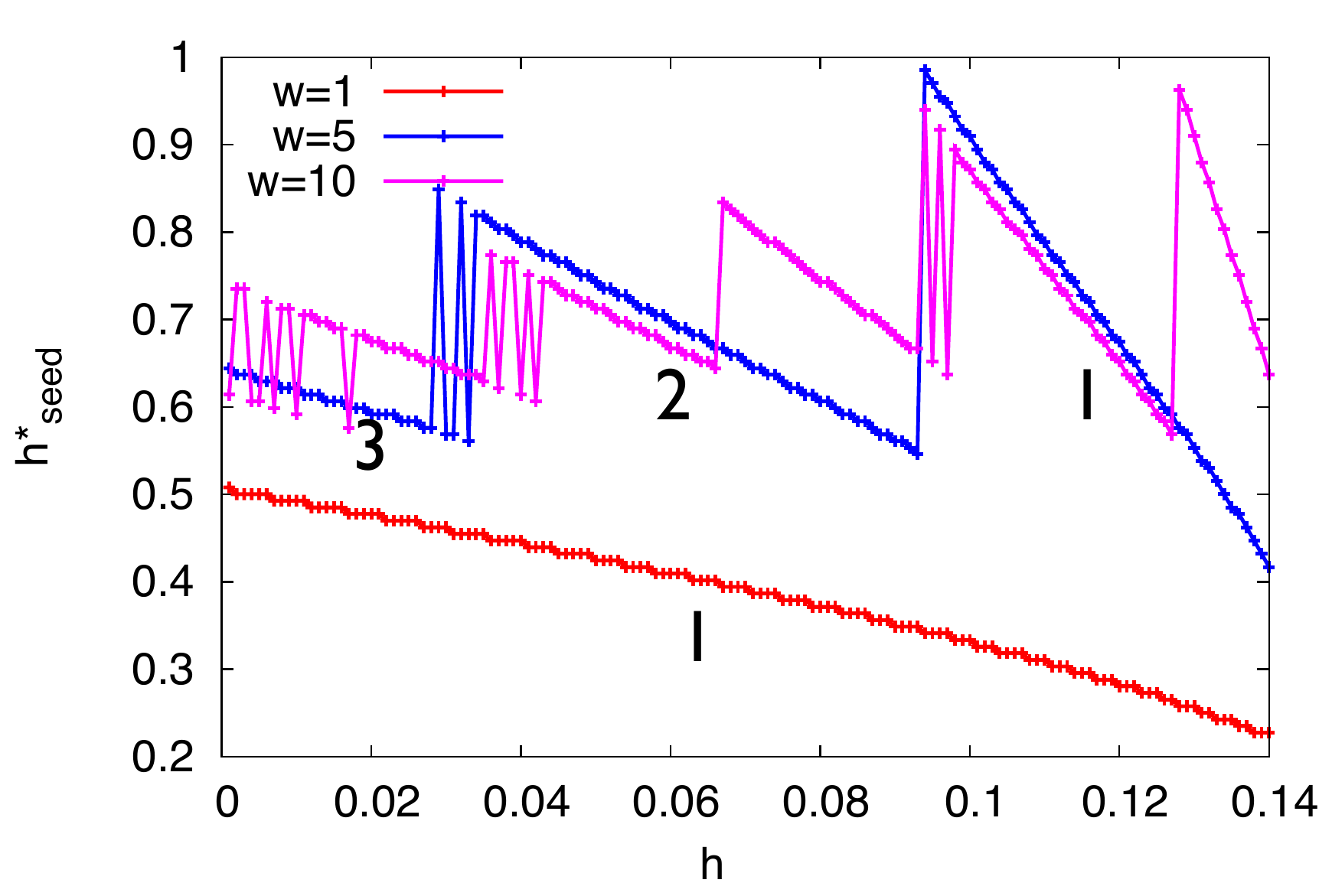}
	\caption{(Color online) With the same parameters as Fig. \ref{fig:mincost}, the three curves represent the seeding field 
	that minimizes the cost as a function of the bulk field. The
        superimposed numbers are the values of the seeding range
        $w_{\rm seed}$
        corresponding to the minimum (shown only for $w=1$ and $w=5$). }
	\label{fig:minseed}
\end{figure}

As can be seen from Fig. \ref{fig:mincost} the minimal total cost is
reached for $w=1$ and an appropriate choices of 
the seeding parameters (see Fig. \ref{fig:minseed}). These two figures
leads us to a perhaps unexpected conclusion that the optimized seeding
condition uses range of interactions $w=1$ with seed size $w_{\rm
  seed}=1$ and sufficiently large $h_{\rm seed}$ (as specified in
Fig. \ref{fig:minseed}).

\section{Conclusions}
\label{sec:concl}

In this paper we have evaluated several properties of the spatially
coupled C-W model. We reached the following interesting conclusions:

\begin{itemize}

\item The speed of propagation of the travelling wave can be increased
by choosing interaction profile that has a large variance.

\item The interaction range $w$ does not really need to be large,
since the negative effects associated with finite $w$ are only visible
at extremely small values of the magnetic field (even for $w = 1$).
Indeed, in practical situations, both the measurement noise and finite
size effects will play an important role, which makes such small
values of the external field--- \textit{i.e.}, such that the effects
of finite $w$ are felt--- unfeasible.

\item In order to minimize the termination cost (or in coding
language, the rate loss) the optimal parameters are: an interaction
range $w=1$, seed size $w_{\rm seed}=1$, and values of the seeding
field $h_{\rm seed}$ summarized in Fig.~\ref{fig:minseed}.

\end{itemize}

In future work we need to clarify whether the above conclusions are
particular to a chain of C-W models, or whether they apply universally
to other spatially coupled systems.

One important aspect that the coupled C-W model is missing are
non-trivial finite size effects. Indeed, the C- W model does not have
any disorder and hence there is a very little difference between the
state evolution and the real evolution in a system of finite size.
This will be different in interesting applications of spatially
coupled systems and needs to be evaluated.

\begin{acknowledgments}
We would like to thank Florent Krzakala for insightful comments and discussions. The research leading to these results was supported by the European Research Council un
der the European Union’s 7th Framework Programme (FP/2007-2013)/ERC
Grant Agreement 307087-SPARCS, and by the Grant DySpaN of "Triangle de la Physique" .
The research of SF has received funding from the European Union, Seventh Framework
Programme FP7-ICT-2009-C under grant agreement n. 265496.
\end{acknowledgments}



\bibliographystyle{unsrt}
\bibliography{refs}

\end{document}